\documentstyle[12pt]{article}

\def\llsymbol#1{\@llsymbol{\@nameuse{c@#1}}}
\def\@llsymbol#1{\ifcase#1\or {}\or {'}\or {''}\or {'''}\or
   {''''}\or {'''''}\or  \else\@ctrerr\fi\relax}

\newcounter{contador}
\newcommand{\letra}{
   \stepcounter{equation}
   \setcounter{contador}{\value{equation}}
   \setcounter{equation}{0}
   \renewcommand{\theequation}{\thecontador.\alph{equation}}}
\newcommand{\antiletra}{
   \renewcommand{\theequation}{\arabic{equation}}
   \setcounter{equation}{\value{contador}}}

\def\acknowledgement{\if@twocolumn
\section*{Acknowledgements}
\else \normalsize
\begin{center}
{\bf Acknowledgements\vspace{-.5em}\vspace{0pt}}
\end{center}
\quotation
\fi}
\def\endacknowledgement{\if@twocolumn\else\endquotation\fi}

\def\I{\leavevmode\hbox{\small1\kern-3.8pt\normalsize1}}
\def\openone{\leavevmode\hbox{\small1\kern-3.3pt\normalsize1}}

\newcommand{\ol}\overline
\newcommand{\wt}\widetilde
\newcommand{\dg}\dagger
\newcommand{\C}{^{\mbox{\scriptsize c}}}

\renewcommand{\and}{\;\;\;\mbox{and}\;\;\;}

\newcommand{\be}{\begin{equation}}
\newcommand{\ee}{\end{equation}}
\newcommand{\bl}{\begin{eqnarray}&}
\newcommand{\el}{&\end{eqnarray}}
\newcommand{\bq}{\begin{eqnarray}}
\newcommand{\eq}{\end{eqnarray}}

\newcommand{\ts}{\textstyle}

\newcommand{\etat}{$\setlength{\unitlength}{.96cm}
                   \begin{picture}(.4,.3)
                   \put(-.43,.37){\line(5,-2){1.25}}
                   \put(-.1,.02){\makebox(0,0){$\eta$}}
                   \put(.55,.21){\makebox(0,0){$t$}}
                   \end{picture}$}

\begin{document}

\title{Some Properties of Charge-Conjugated Spinors in $D$ dimensions}

\author{M. A. De Andrade\thanks{Internet e-mail: marco@cbpfsu1.cat.cbpf.br}\\
Centro Brasileiro de Pesquisas F\'\i sicas (CBPF) \\
Departamento de Teoria de Campos e Part\'\i culas (DCP)\\
Rua Dr. Xavier Sigaud, 150 - Urca \\
22290-180 - Rio de Janeiro - RJ - Brasil.}

\date{}

\maketitle
\begin{abstract}
Spinors for an arbitrary Minkowski space with signature ($t$, $s$) are
reassessed in connection with $D$-dimensional free Dirac action. The
possibility of writing down kinetic and mass terms for charge-conjugated
spinors is discussed in terms of the number of time-like directions of the
space-time.
The results found out here are commented in the light of early works on the
subject.
\end{abstract}

Spinors in arbitrary Minkowski-like spaces with signature ($t$, $s$), M$^{t,s}$
, have been carefully investigated in the papers of Refs.
\cite{Wett,Kugo,Sohn}. There are several ways of imposing suitable conditions
in order to restrict the number of degrees of freedom carried by the spinors,
according to the particular space-time one is considering.

Usually, one discusses the kinematical aspects of spinors by dealing with the
Dirac equation in the space-time M$^{t,s}$.
A collection of interesting results can be found in a very complete paper by
Kugo and Townsend \cite{Kugo}. With the requirements that the Dirac equation be
derivable from a Lagrangian and that the spinors be described by anticommuting
components, we adopt here the viewpoint of discussing the kinematical
properties of spinors in $D$ dimensions through a free Dirac-like action,
rather than in terms of the free Dirac equation itself. In particular, we find
a number of peculiar results regarding the propagation of charge-conjugated
spinors, so that Majorana and Majorana-Weyl spinors are directly concerned by
our discussion.

We shall now present our analysis and the main results shall be discussed in
our general conclusions. The considered metric for the space-time M$^{t,s}$ is
the generalized Minkowski one:
\be
\eta^{\mu\nu}=\mbox{diagonal}(\underbrace{+,+,\cdots,+}_{t-\mbox{\scriptsize
times}}\;,\;\underbrace{-,-,\cdots,-}
_{s-\mbox{\scriptsize times}}),\;\;\; \mu,\; \nu=(1,\cdots,D)\;.
\ee
The Dirac $\gamma$-matrices for a space-time of dimension $D$ = $t$+$s$ can be
represented by $2^{[D/2]}\times2^{[D/2]}$ complex matrices (the bracket $[D/2]$
denotes the integral part of $D/2$) that satisfy the Clifford algebra:
\be
\{\gamma^\mu,\gamma^\nu\}=2\eta^{\mu\nu}\I\;,
\ee
where $\I$ is the $2^{[D/2]}\times2^{[D/2]}$ identity matrix. The  $\gamma$'s
which satisfy $\gamma^{\mu2}$ = $\I$ and $\gamma^{\mu2} = -\I$ will be
respectively called $t$-like and $s$-like. The matrices $\gamma^\mu$,
$\gamma^{\mu\dg}$, $\gamma^{\mu*}$ and $\gamma^{\mu T}$, up to constant
factors, yield equivalent  representations of the Clifford algebra; thus, for a
given $t$:
\letra
\bq
\gamma^{\mu\dg}&=&-(-1)^tA\gamma^\mu A^{-1}\;,         \label{gmdag} \\
\gamma^{\mu *} &=&\eta B\gamma^\mu B^{-1}\;,           \label{gmstar}\\
\gamma^{\mu T} &=&-\eta(-1)^tC\gamma^\mu C^{-1}\;;     \label{gmtran}
\eq
\antiletra
the $t$-like and $s$-like $\gamma$-matrices can always be chosen Hermitean and
anti-Hermitean, respectively. It can be seen, by using the Schur's lemma, that
such a choice amounts to taking $A$, $B$ and $C$ as unitary matrices. The {\em
unitarity} of $A$, $B$ and $C$ also plays a central role in the construction of
a real action for the spinors and their corresponding charge-conjugated
partners. The unitary matrix that satisfies (\ref{gmdag}) is given by
\be
A=\gamma^1\cdots\gamma^t\;.
\ee
As we shall see, the $\eta$-factor in (\ref{gmstar}) will have its value
($\pm$1) fixed, for a given $t$, through the existence, or not, of the action
for the charge-conjugated spinor. The factor in (\ref{gmtran}) follows from the
consistency with the previous one. So, the matrices $A$, $B$ and $C$ exhibit
the properties \cite{Kugo}:
\letra
\bq
B^T&=&\xi CA^{-1}\;,\;\;\xi\mbox{ being a phase factor,}
\label{Cdef}\\
&&\nonumber\\
A^{-1}&=&(-1)^{t(t-1)/2}A\;,                     \label{invA}\\
B^T&=&\varepsilon B\;,\;\;\;\varepsilon=\pm1\;,  \label{trB}\\
C^T&=&\varepsilon\eta^t(-1)^{t(t-1)/2}C\;,\\
&&\nonumber\\
A^*&=&\eta^tBAB^{-1}\;,\\
A^T&=&\eta^tCA^{-1}C^{-1}\;,\\
\xi C&=&\eta^tB^T(\xi C)^*B\;.
\eq
\antiletra

The $\varepsilon$-factor appearing in (\ref{trB}), for dimensions $D$ =
($t$+$s$) and ($D$+1) (with $D$ even), is shown to be given by
\cite{Kugo,Glio,Sche}
\be
\varepsilon =  \cos\frac\pi4(s-t)-\eta\sin\frac\pi4(s-t)=\pm1.  \label{epsilon}
\ee

Spinors are objects that, under the action of the group SO$_0$($t$,$s$),
transform as
\be
\Psi\rightarrow\Psi^\prime=e^{\frac12\omega_{\kappa\lambda}
\Sigma^{\kappa\lambda}}\Psi\;,\;\;\;\Sigma^{\kappa\lambda}={\ts\frac14}
[\gamma^\kappa,\gamma^\lambda].                \label{espinor}
\ee
A conjugated spinor, $\ol\Psi$, can be defined in a way such that the bilinear
$\ol\Psi\Psi$ be a scalar under the SO$_0$($t,s$) group transformation. Using
(\ref{gmdag}), it can be readily seen that, under such a transformation,
\be
\Psi^\dg A\rightarrow\Psi^{\prime\dg}A=\Psi^\dg A
e^{-\frac12\omega_{\kappa\lambda}\Sigma^{\kappa\lambda}}.
\ee
Therefore, $\ol\Psi$ can be identified with
$\ol\Psi$ = $\Psi^\dg A$. Using the relation
\be
e^{-\frac12\omega_{\kappa\lambda}\Sigma^{\kappa\lambda}}\gamma^\mu
e^{\frac12\omega_{\rho\sigma}\Sigma^{\rho\sigma}}
=\Lambda^\mu_{\;\;\;\nu}\gamma^\nu\;,
\ee
where $\Lambda^\mu_{\;\;\;\nu}$ is the transformation matrix of the
SO$_0$($t,s$) group for vectors, it can be found that $\ol\Psi\gamma^\mu\Psi$
transforms like a vector
under SO$_0$($t,s$) transformations.

The free Dirac action, $\cal A$, is a real functional and a scalar under
SO$_0$($t$,$s$). It must carry infor\be
{\cal A}=\int d^Dx{\cal L}(\Psi,\partial_\mu\Psi).
\ee
The Dirac Lagrangian, $\cal L$, up to a ``surface term'' (ST), has the form
\be
{\cal L}(\Psi,\partial_\mu\Psi)=\alpha\ol\Psi\gamma^\mu\partial_\mu\Psi+\beta
m\ol\Psi\Psi\;,                               \label{L}
\ee
where the parameter $m$ is real and $\alpha$ and $\beta$ are factors that may
assume the values ($\pm1$), ($\pm i$), depending on the number of time-like
directions. The determination of $\alpha$ and $\beta$, for a given $t$, can be
achieved through the Hermiticity of $\cal L$. Using (\ref{gmdag}), and
enforcing the Grassmannian character of the spinors, it can be found, up to a
ST, that
\be
{\cal
L}^\dg=\alpha^*(-1)^{t(t+1)/2}\,\ol\Psi\gamma^\mu\partial_\mu\Psi+
\beta^*(-1)^{t(t-1)/2}\,m\ol\Psi\Psi\;.    \label{Ldag}
\ee
In order that $\cal A$ be real, one must have:
\letra
\bq
\alpha^*(-1)^{t(t+1)/2}&=&\alpha\;,          \label{alpha}\\
\beta^* (-1)^{t(t-1)/2}&=&\beta \;.          \label{beta}
\eq
\antiletra
With the aid of these equations, we can build up Table 1, where the values of
$\alpha$ and $\beta$ are displayed for different numbers of time-like
directions.

\vspace{.5cm}

\begin{center}
\begin{tabular}{||c||c|c|c|c||}  \hline\hline
$t$ & 0 mod 4 & 1 mod 4 & 2 mod 4 & 3 mod 4 \\ \hline\hline
$\alpha$ & 1 & $i$ & $i$ & 1 \\ \hline
$\beta$ & 1 & 1 & $i$ & $i$ \\\hline\hline
\end{tabular}
\end{center}
\vspace{.3cm}
\centerline{Table 1: Conditions for the reality of the Dirac action $\cal A$.}

\vspace{.5cm}

Taking now the complex conjugate of $\cal L$, and using (\ref{gmstar}), it can
be found that
\be
{\cal
L}^*=-\alpha^*\eta^{t+1}\,\ol{\Psi\C}\gamma^\mu\partial_\mu
\Psi\C-\beta^*\eta^t\,m\ol{\Psi\C}\Psi\C\;,         \label{Lstar}
\ee
where
\be
\Psi\C\equiv B^{-1}\Psi^*\and\ol{\Psi\C}=(\Psi\C)^\dg A\;.
\ee
The transformation $\Psi\rightarrow\Psi\C$ is the so-called charge conjugation
operation. Next, assuming the Hermiticity of $\cal L$, and upon use of the
equations (\ref{alpha}) and (\ref{beta}), it results that
\be
{\cal L}^*={\cal L}^T={\cal L}^\prime(\Psi\C,\partial_\mu\Psi\C)=
-\alpha\eta^{t+1}(-1)^{t(t+1)/2}\,\ol{\Psi\C}\gamma^\mu\partial_\mu\Psi\C-\beta
\eta^t(-1)^{t(t-1)/2}\,m\ol{\Psi\C}\Psi\C\;.
                                               \label{LCC}
\ee
On the other hand, taking the transposition of $\cal L$, and using
(\ref{gmtran}), it can be found, up to a ST, that
\be
{\cal
L}^T=-\alpha\eta^{t+1}(-1)^{t(t+1)/2}\,\ol{\Psi^\prime}\gamma^\mu
\partial_\mu\Psi^\prime-\beta \eta^t(-1)^{t(t-1)/2}\,m\ol{\Psi^\prime}
\Psi^\prime\;,
                                               \label{Ltransp}
\ee
where
\be
\Psi^\prime=\eta^t(-1)^{t(t-1)/2}\,C^{-1}\ol\Psi^T \and
\ol{\Psi^\prime}=\Psi^TC\;.
\ee
{}From this result, one may conclude, by comparing with (\ref{LCC}), that
$\Psi^\prime\propto\Psi\C$. It can be quickly computed that
$\Psi\C$ = $\xi^*\varepsilon\Psi^\prime$ ($\xi$ is the phase factor in
(\ref{Cdef})), so that
\be
\Psi\C=\xi^*C^*\ol\Psi^T\;.
\ee

A Majorana spinor is defined by the condition $\Psi\C$ = $\Psi$, or
equivalently
\be
\Psi=B^{-1}\Psi^*=\xi^*C^*\ol\Psi^T\;.              \label{Maj}
\ee
Note that, in the so-called Majorana representation ($B$ = $\I$), a Majorana
spinor is always {\em real}. It is worthwhile to mention that Majorana spinors
can only be defined for M$^{t,s}$ such that $\varepsilon$ = 1, as it will be
verified in the sequel. Inverting (\ref{Maj}) for $\Psi^*$, it can be obtained
that
\be
\Psi^*=B\Psi\;.                              \label{conta}
\ee
Complex conjugation of the previous equation yields
\be
\Psi=B^*\Psi^*\;.                            \label{conta2}
\ee
Next, plugging (\ref{conta}) into (\ref{conta2}), it can be found that $B^*B$ =
$\I$, so that, from (\ref{trB}), it results $\varepsilon$ = 1.\footnote{In the
case $\varepsilon$=$-$1, one can impose, for example, the so-called
SU(2)-reality condition and define SU(2)-Majorana spinors\cite{Kugo}.}

The free Lagrangian for  the Majorana spinor $\Psi$ takes the form
\be
{\cal L}=\alpha\Psi^T\xi C\gamma^\mu\partial_\mu\Psi+\beta \;m\Psi^T\xi
C\Psi\;.
\ee

Before our final considerations on the $\eta$-factor, let us discuss also the
case of Majorana-Weyl spinors. For even $D$, we can  decompose
$\Sigma^{\kappa\lambda}$ in generators  $\Sigma^{\kappa\lambda}_+$ and
$\Sigma^{\kappa\lambda}_-$ of independent transformations, which determine
complementary sectors in the $\Psi$ space ($\Psi_+$ and $\Psi_-$ chiral
subspaces respectively). In the case ($s$$-$$t$) = 0 mod 4,
$\Sigma_\pm^{\kappa\lambda}$ are real in the Majorana representation
\cite{Kugo}. A Majorana spinor in the Majorana representation is real, such
that its sector $\Psi_+$, when transformed via $\Sigma_+^{\kappa\lambda}$, will
always be {\em real}. Analogously, $\Psi_-$, when transformed via
$\Sigma_-^{\kappa\lambda}$, will always be {\em real}: that is, the fundamental
property of Majorana spinors in the Majorana representation (the reality) is
preserved in each chirality sector. This guarantees that we can expect that the
property $\Psi_\pm^*$ = $B\Psi_\pm$, under transformation via the generators
$\Sigma_\pm^{\kappa\lambda}$, will be preserved in each corresponding sector,
for any chosen representation.
$\Psi_+$ and $\Psi_-$, with the properties shown above are known as
Majorana-Weyl spinors.
We are now ready to state our general results on the $\eta$-factor, and we
shall quote some interesting properties on the dynamics of the
charge-conjugated spinors.

{}From the invariance of the Lagrangian under charge conjugation operation, it
follows, by virtue of  (\ref{L}) and (\ref{LCC}), that
\letra
\bq
\eta^{t+1}(-1)^{t(t+1)/2}&=&-1\;,            \label{eta1}\\
\eta^t(-1)^{t(t-1)/2}    &=&-1\;.            \label{eta2}
\eq
\antiletra
Relations (\ref{eta1}) and (\ref{eta2}) do {\it not} show up from an analysis
based directly on the Dirac equation; they govern respectively the existence of
the {\em kinetic term} and {\em mass term} of the action for the
charge-conjugated spinor (and consequently for the Majorana spinor).

We present now some general conclusions concerning the existence of the kinetic
and mass terms for the charge-conjugated spinors:

a) For even $t$, from (\ref{eta1}) it results that kinetic terms can always be
written down, since $\eta$ = $(-1)^{\frac t2+1}$. In this case, (\ref{eta2})
turns out to be a consistency condition, from which it follows that a mass term
is only possible for $t$ = 2 mod 4, such that $\eta$ = 1.

b) On the other hand, for odd $t$, (\ref{eta1}) becomes a consistency
condition, and then it rules out the existence of the kinetic action for the
charge-conjugated spinor in the case $t$ = 3 mod 4. So, a kinetic term is
possible only for $t$ = 1 mod 4, so that the existence of mass for these
spinors determines $\eta = -1$.

c) We therefore conclude that for $t$ = 1, 2 mod 4, we can always write down
actions with both kinetic and mass terms for the charge-conjugated spinor. For
$t$ = 0 mod 4, the charge-conjugated spinor cannot be massive; whereas for $t$
= 3 mod 4, it cannot be dynamical (no kinetic term).

d) When the value of $\eta$ is such that it prohibits mass terms, the condition
$\Psi\C$ = $\Psi$ defines the so-called pseudo-Majorana spinor. According to
the paper of \cite{Kugo}, this type of spinor appears indeed to be massless.

The results mentioned above are all summarized in Table 2.

\vspace{.5cm}

\begin{center}
\begin{tabular}{||c||c|c|c|c||} \hline \hline
\etat & 0 mod 4 & 1 mod 4 & 2 mod 4 & 3 mod 4 \\
\hline\hline
1 &  --- & pseudo-Majorana & Majorana & --- \\ \hline
$\;-1\;$ & pseudo-Majorana & Majorana & --- & --- \\\hline\hline
\end{tabular}
\end{center}
\vspace{.3cm}
\noindent
Table 2: Types of self-conjugated spinors ($\varepsilon$=1) as a function of
$\eta$ and the number, $t$, of time-like directions. --- means that no kinetic
term can be written down for charged-conjugated spinors with the pair of values
($t$,$\eta$).

To conclude this letter, we should stress that our results on the possibility
of writing down kinetic and mass Lagrangians for charge-conjugated spinors rely
on our assumption that spinors be always taken to be Grassmann-valued, no
matter what the space-time signature is. For non-usual signatures with
$t$$\geq$2, instead of giving up a kinetic term for charge-conjugated spinors
in the cases indicated in Table 2, one could perhaps decide to work with
commuting spinors, for which a kinetic term becomes non-trivial. Nevertheless,
this is not the viewpoint we take here: we assume spinors to be anticommuting
in any case, for we have in mind to formulate supersymmetric field theories in
space-times with arbitrary signatures, and so these spinors may well be
identified with supersymmetry charge generators and fermionic matter that, as
it is well-known, should always exhibit an anticommuting character.

\vspace{.5cm}

\begin{acknowledgement}
The author express his gratitude to Dr. J. A. Helay\"el-Neto and O. M. Del Cima
for patient and helpful discussions. Thanks are also due our colleagues at
CBPF-DCP. CNPq-Brazil is acknowledged for invaluable financial help.
\end{acknowledgement}

\label{refe}

\end{document}